\documentclass{article}
\usepackage{spconf,amsmath,amssymb,graphicx,hyperref,booktabs, multirow}
\usepackage[font=footnotesize,labelfont=bf]{caption}
\usepackage[table]{xcolor}
\usepackage{array}
\newcolumntype{C}[1]{>{\centering\arraybackslash}p{#1}}
\usepackage{makecell}

\title{Sound Separation and Classification with Object and Semantic Guidance}
\name{Younghoo Kwon$^{1}$ and Jung-Woo Choi$^{1}$\sthanks{Corresponding author.}}
\address{$^{1}$School of Electrical Engineering, KAIST, Daejeon, Korea \\ \{k0hoo, jwoo\}@kaist.ac.kr}
\begin{document}
\ninept
\maketitle
\begin{abstract}
The spatial semantic segmentation task focuses on separating and classifying sound objects from multichannel signals. To achieve two different goals, conventional methods fine-tune a large classification model cascaded with the separation model and inject classified labels as separation clues for the next iteration step. However, such integration is not ideal, in that fine-tuning over a smaller dataset loses the diversity of large classification models, features from the source separation model are different from the inputs of the pretrained classifier, and injected one-hot class labels lack semantic depth, often leading to error propagation. To resolve these issues, we propose a Dual-Path Classifier (DPC) architecture that combines object features from a source separation model with semantic representations acquired from a pretrained classification model without fine-tuning. We also introduce a Semantic Clue Encoder (SCE) that enriches the semantic depth of injected clues. Our system achieves a state-of-the-art 11.19 dB CA-SDRi and enhanced semantic fidelity on the DCASE 2025 task4 evaluation set, surpassing the top-rank performance of 11.00 dB. These results highlight the effectiveness of integrating separator-derived features and rich semantic clues.

\end{abstract}
\section{Introduction}
\label{sec:intro}

Comprehending auditory scenes is crucial in machine listening, as it allows systems to interpret and engage with complex soundscapes. To fulfill this aim, it is necessary for individual sound sources to be separated and categorized. While the performance of sound separation improves with the advancements of the models~\cite{10738447, 10094992}, conventional separation models are limited by the maximum number of sources they can process. To overcome this limitation, researchers have developed methods that can pinpoint and isolate a specific sound from the complex sound scene by using guiding information. For example, the system can be told what to listen for using a class label~\cite{10094573}, a similar audio sample~\cite{9944179, 10888410}, or a timestamp~\cite{10447525}. Meanwhile, classifiers~\cite {task4first, task4second, 10446159} have been developed to categorize each sound source using the features from the self-supervised models~\cite{fasst, atst, m2d} pretrained on a large-scale dataset~\cite{gemmeke2017audio}. In pursuit of understanding the auditory scene, one strategy for joint separation and classification is first to perform multi-label classification on the mixture, using the resulting class predictions as clues for target sound extraction~\cite{9053921, 10447061}. Another strategy, embodied by DeFT-Mamba~\cite{deftmamba}, is first to separate each source from the mixture and then classify them.

\begin{figure}
    \centering
    \includegraphics[width=1\linewidth]{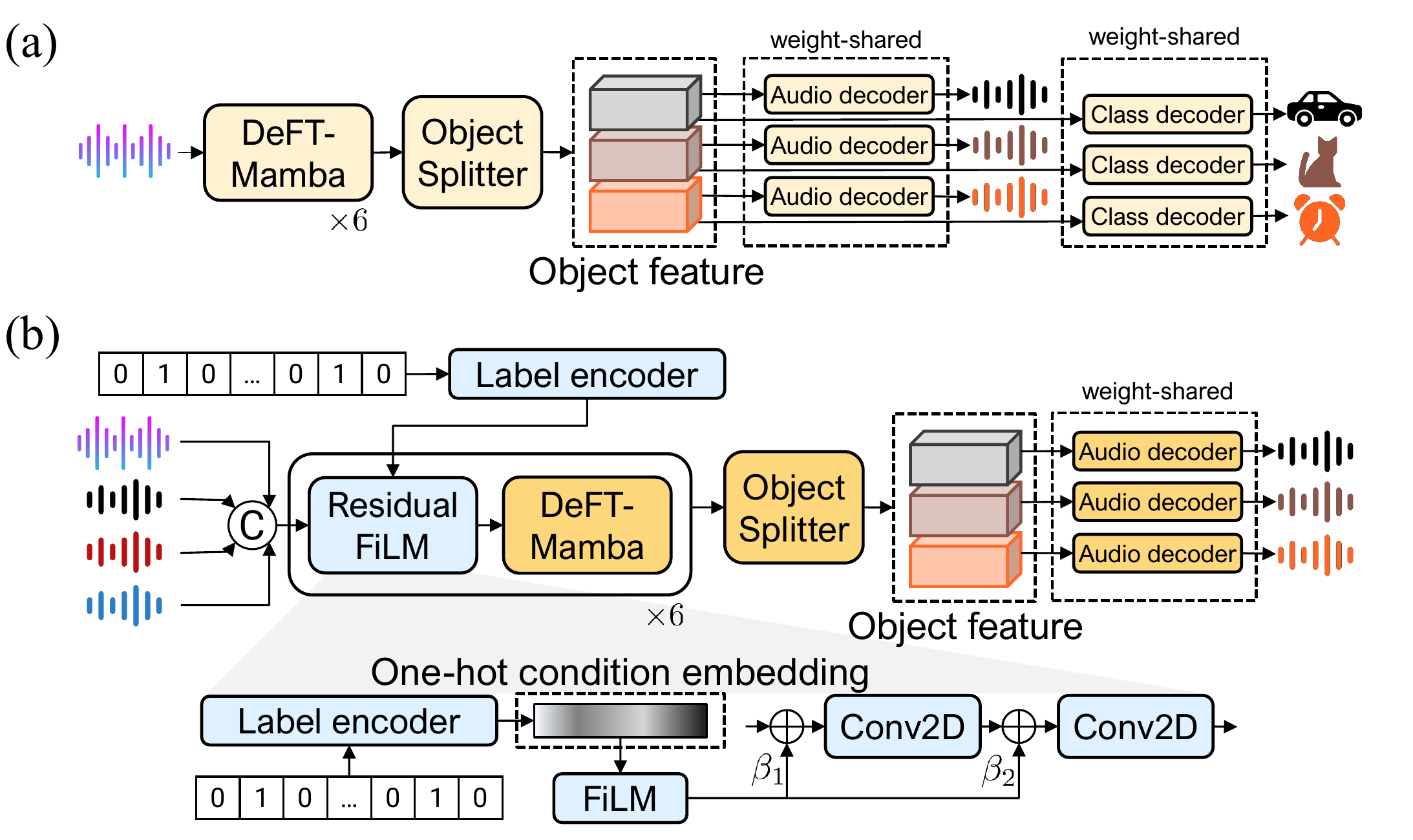}
    \caption{Architectures utilized for the top-ranked framework in DCASE 2025 task4. (a) Universal separation model (DeFT-Mamba-USS) for Stage 1. (b) Target signal extraction model (DeFT-Mamba-TSE) and its Residual FiLM for Stage 2 and 3.}
    \label{fig:DM}
    \vspace{-8pt}
\end{figure}

The DCASE 2025 task4 challenge, known as Spatial Semantic Segmentation of Sound Scenes (S5)~\cite{yasuda2025, nguyen2025}, has been designed to reflect the growing focus of research. The challenge focuses on separating and classifying foreground sound sources from complex multichannel mixtures, which also include interfering sounds and background noises. Each mixture can contain up to three foreground sources, and each separated foreground signal is categorized into one of the 18 designated classes. 
\begin{figure*}
    \centering
    \includegraphics[width=0.85\linewidth]{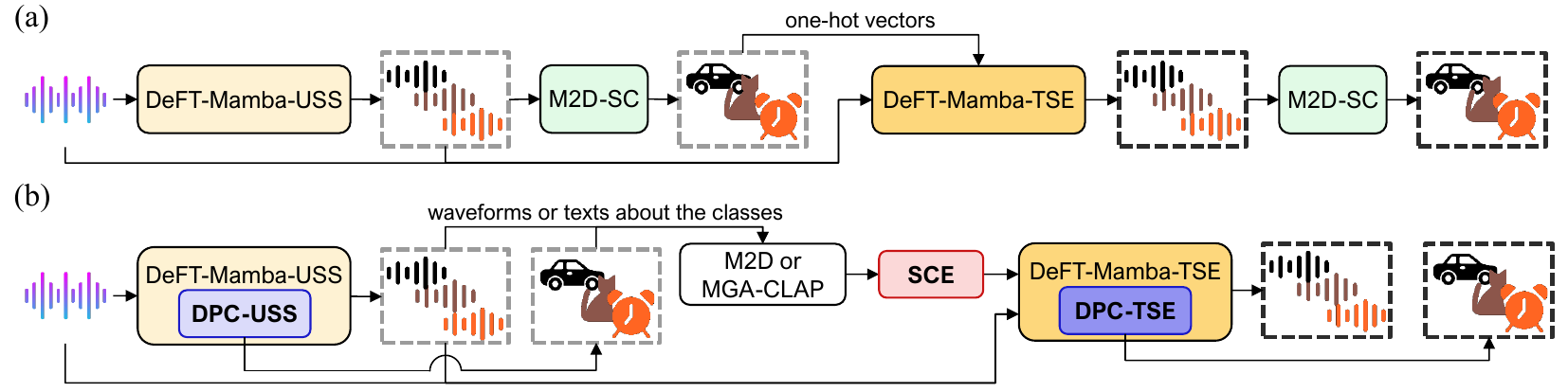}
    \caption{The comparison of the proposed framework with the architecture used for the DCASE 2025 task4 challenge. (a) The framework used for challenge submission with M2D-SC. (b) The proposed framework incorporated with the Dual-Path Classifier (DPC) and the Semantic Clue Encoder (SCE).}
    \label{fig:framework}
    \vspace{-8pt}
\end{figure*}

Our previous work, achieved the 1st rank in the challenge~\cite{techreport}, employs a three-stage framework. 
Stage 1 (\textbf{clue derivation}) separates object features without any clues, using DeFT-Mamba for Universal Sound Separation (DM-USS). As illustrated in Fig.~\ref{fig:DM} (a), the separated object features are decoded into the waveforms and class estimations by the audio decoder and the class decoder, respectively, whose weights are shared across all object features. 

In stage 2 (\textbf{guided extraction}), Target Sound Extraction (TSE) is performed using the clues developed in Stage 1. As illustrated in Fig.~\ref{fig:DM} (b), DeFT-Mamba for TSE (DM-TSE). The mixtures and three separated waveforms (enrollment clues) are concatenated along the channel axis~\cite{11027436}. Additionally, one-hot class labels are estimated from a large classification model, M2D~\cite{m2d} for Single-label Classification (M2D-SC). As depicted in Fig.~\ref{fig:DM} (b), one-hot vectors are fed into a label encoder to generate a one-hot condition embedding. This embedding is then injected as a class clue via the Residual FiLM~\cite{perez2018film, kong2023universal} module before each DeFT-Mamba block to guide the extractor. 
Stage 3 (\textbf{iterative refinement}) follows the same procedure as Stage 2. The extracted waveforms are fed into DM-TSE as enrollment clues, along with class clues represented as one-hot vectors from M2D-SC, forming an iterative loop to refine estimations.

The current design exhibits several limitations due to the coarse combination with the existing classification model. (1) The M2D-SC is based on the M2D pretrained over the large corpus~\cite{gemmeke2017audio}, but fine-tuning it on the smaller S5 dataset~\cite{yasuda2025} compresses the pretrained diversity~\cite{Yamaguchi_2024_CVPR, Perera_2024_CVPR}. 
(2) Separated object features are converted into Mel-spectrograms to make them compatible with M2D. However, the magnitude-only projections with spectral merging lead to a significant loss of crucial phase and fine-grained frequency information. (3) The one-hot condition embedding based on one-hot vectors fails to encode rich semantic information. (4) The class-based guidance for the TSE model depends solely on the predicted label so that an early misclassification can steer DM-TSE in the wrong direction throughout the extraction stages.
Therefore, we need a novel classification architecture that
can leverage the knowledge obtained from the pretraining and inject semantic clues with sufficient depth.
\begin{figure}
    \centering
    \includegraphics[width=0.95\linewidth]{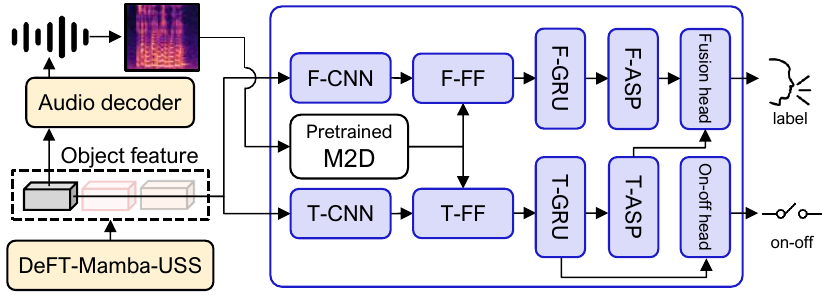}
    \caption{The proposed pipeline of the Dual-Path Classifier for the object features from DeFT-Mamba-USS (DPC-USS).}
    \label{fig:classifier}
    \vspace{-14pt}
\end{figure}

We propose a Dual-Path Classifier (DPC) to address the lack of diversity in the M2D-SC and the informative insufficiency of the Mel-spectrogram. We adopt the lightweight CRNN instead of fine-tuning M2D to preserve the diversity of the features from M2D. The object features of the separators are fed into the CRNN instead of the Mel-spectrogram. Unlike Mel-spectrogram, these features retain crucial phase and fine-grained frequency information, providing a richer input for the classifier. Additionally, the DPC adopts two branches to capture distinct temporal and spectral information. Within each of these specialized paths, the CRNN takes advantage of features from the pretrained M2D by integrating them with the object features through a dedicated fusion block within each branch.

To overcome the static nature of the one-hot condition embedding and the critical flow of misclassification, we propose a Semantic Clue Encoder (SCE). The SCE solves this issue by supplementing the conventional one-hot condition embedding with richer embeddings from models trained with self-supervised methods~\cite{m2d} or contrastive learning~\cite{mgaclap}. The injection of these additional semantic clues through the SCE makes the DM-TSE less critically dependent on the label estimation of the previous stage, thereby enhancing the overall robustness of the system. Eventually, our system achieves a final CA-SDRi of 11.19 dB on the evaluation dataset and 16.60 dB on the test dataset~\cite{nguyen2025}, corresponding to gains of +0.19 dB and +1.66 dB, respectively, over our previous submission.

\section{Methodology and architecture}
\label{sec:method}

As shown in Fig.~\ref{fig:framework}, the proposed architecture applies DPC and enhances the DM-TSE model with the SCE using the semantic condition embeddings. Two individual instances of the DPC are trained: one for DM-USS (DPC-USS) and another for DM-TSE (DPC-TSE).

\subsection{Dual-Path Classifier (DPC)}
\label{subsec:class_decoder}
The proposed DPC is detailed in Fig.~\ref{fig:classifier}. The core of DPC is a dual-path CRNN that operates on the object features of the separators. The CRNN consists of symmetric temporal and frequency paths (T/F-paths), each containing a sequence of a CNN, a Feature Fusion (FF) block, a GRU~\cite{gru}, and an ASP~\cite{asp} layer. The feature from each path is mixed to produce logits in the fusion head. Additionally, the on-off head determines whether the separated source is the silent signal or not.

\noindent\textbf{T/F-CNN:}
At first, a stack of 2D convolutions and pooling layers compresses the object feature $\mathbf{X}\in\mathbb{R}^{D\times F\times T}$ along the time and frequency axes. The output of CNN in the temporal path is $\mathbf{X}_t\in\mathbb{R}^{D\times F_t\times T_t}$, while the frequency path employs a final point-wise convolution that doubles the channel dimension, resulting in $\mathbf{X}_f\in\mathbb{R}^{2D\times F_f\times T_f}$. 
The features are then reshaped for sequential processing. The temporal feature $\mathbf{X}_t$ is flattened to preserve the time axis, yielding $X_t\in\mathbb{R}^{T_t\times DF_{t}}$, while the frequency feature $\mathbf{X}_f$ averages out its time dimension regarding its frequency as a sequence dimension, resulting $\mathbf{X}_{f}\in\mathbb{R}^{F_f\times 2D}$.

\noindent\textbf{Feature Fusion (FF) block and GRU:}
The FF block aggregates the features of M2D and CNN in each path. A feature $\mathbf{X}_{\mathrm{M2D}}\in\mathbb{R}^{G\times F_{\mathrm{M2D}}\times T_{\mathrm{M2D}}}$ is extracted from the M2D to leverage knowledge from a pretrained model. To prepare for fusion, the $\mathbf{X}_{\mathrm{M2D}}$ is reshaped differently for each path to match the dimensions of the corresponding CNN feature. For the temporal path, $\mathbf{X}_{\mathrm{M2D}}$ is first average-pooled along the frequency axis and then upsampled via bilinear interpolation along the time axis to match the sequence length of the CNN feature, resulting $\mathbf{X}_{\mathrm{M2D}, T}\in\mathbb{R}^{G\times T_t}$. For the frequency path, a symmetric operation is performed to produce $\mathbf{X}_{\mathrm{M2D}, F}\in\mathbb{R}^{G\times F_f}$. Each aligned M2D feature is then concatenated with its corresponding sequence feature from the CNN along the embedding dimension. These concatenated features are then passed through the FF block, which is the linear layer. This unified feature sequence is then fed into a bidirectional GRU, which models long-range contextual dependencies. 

\noindent\textbf{ASP and Fusion head:}
The ASP layer converts the sequential feature of each GRU into a fixed-size embedding vector. Specifically, each branch contains its own ASP layer that processes the respective GRU output, creating two distinct embedding vectors. The temporal and spectral embeddings are concatenated and passed through a fusion head to produce the logits. The fusion head consists of a linear layer.

\noindent\textbf{On-off prediction:}
The temporal path also handles on-off (silence) prediction using the on-off head on the sequential output of GRU in the temporal path. This head produces frame-wise activity probabilities with a linear layer and sigmoid function, and the source is classified as `on' if the maximum probability across frames exceeds a threshold of 0.5.

\subsection{Semantic Clue Encoder (SCE)}
\label{subsec:semantic_clue}
The original framework~\cite{techreport, dcaseworkshop} conditions the DM-TSE using only estimated labels when conditioning through the Residual FiLM. Specifically, the one-hot vectors corresponding to the foreground sources are concatenated into a single vector. This vector is then passed through a label encoder to generate the one-hot condition embedding. It is then mapped to the modulation parameters $\beta_1$ and $\beta_2$ by the FiLM layer. These parameters modulate the input features of Conv2D layers, as illustrated in Fig.~\ref{fig:DM} (b).

\begin{figure}
    \centering
    \includegraphics[width=0.75\linewidth]{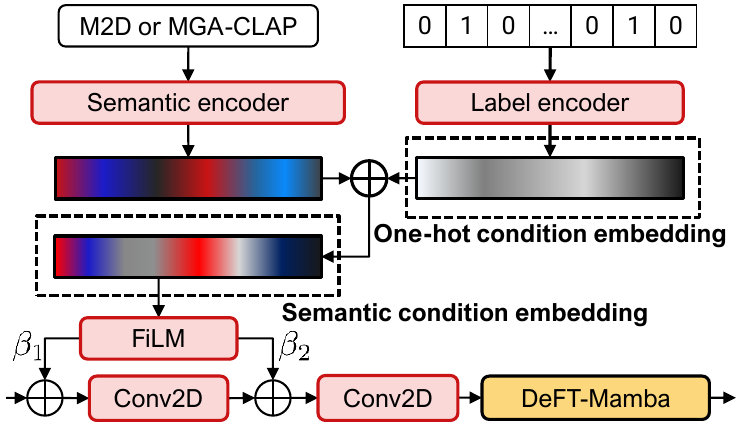}
    \caption{The structure of the Semantic Condition Encoder (SCE) with the definition of the detailed modules and the condition embeddings.}
    \label{fig:semantic}
    \vspace{-8pt}
\end{figure}

As illustrated in Fig.~\ref{fig:semantic}, the SCE denotes the conditioning module extended with semantic embeddings on top of the original label-based condition. Concretely, the SCE includes not only the label encoder that processes one-hot vectors but also a semantic encoder that transforms semantic embeddings extracted from pretrained models such as M2D or MGA-CLAP~\cite{mgaclap}. The condition embeddings produced by these two encoders are then added to form a semantic condition embedding, which is subsequently delivered to the FiLM to generate the modulation parameters, $\beta_1$ and $\beta_2$.
\begin{table*}[t]
\caption{Ablation study on the DPC and SCE across each stage. MGA-CLAP-text and MGA-CLAP-audio indicate that the embedding is generated from the text and audio encoder of MGA-CLAP, respectively. The gray-colored rows represent the performance of the previous model~\cite{techreport}. The hyphen in the `Semantic embedding' column denotes conditioning solely on the one-hot condition embedding without any semantic information. CA-SDRi in [dB] and $\mathrm{Acc}_{\mathrm{mix}}$ in [\%].}
\label{tab:main}
\centering
\setlength{\tabcolsep}{2pt}
{\scriptsize
\begin{tabular}{@{}ccc*{8}{C{1.5cm}}@{}}
\toprule
\multirow[]{2}{*}{Stage} & \multirow[]{2}{*}{Classifier} & \multirow[]{2}{*}{\makecell{Semantic \\ embedding}} & \multicolumn{4}{c}{Eval dataset} & \multicolumn{4}{c}{Test dataset} \\
\cmidrule(lr){4-7} \cmidrule(lr){8-11}
& & & CA-SDRi $\uparrow$ & $\mathrm{Acc}_{\mathrm{mix}}$ $\uparrow$ & PESQ $\uparrow$ & CLAP-score $\uparrow$ & CA-SDRi $\uparrow$ & $\mathrm{Acc}_{\mathrm{mix}}$ $\uparrow$ & PESQ $\uparrow$ & CLAP-score $\uparrow$ \\

\midrule

\rowcolor{gray!15} \cellcolor{white} \multirow[]{2}{*}{1} & M2D-SC & -
  & 8.47 & 52.22 & 2.38 & 0.757 & 12.70 & 59.07 & 2.53 & 0.764 \\
\cmidrule{2-11}
& DPC & -
  & 8.18 & 52.04 & 2.36 & 0.757 & 12.81 & 69.20 & 2.52 & 0.767 \\

\midrule

\rowcolor{gray!15} \cellcolor{white} \multirow[]{5}{*}{2} & M2D-SC & -
  & 10.08 & 54.01 & 2.88 & 0.800 & 14.70 & 61.26 & 3.10 & 0.833 \\
\cmidrule{2-11}
& \multirow[]{4}{*}{DPC} & -
  & 10.86 & 57.78 & 2.91 & 0.800 & 16.21 & 72.13 & 3.13 & 0.837 \\
&  & MGA-CLAP-text
  & 10.97 & 58.27 & 2.91 & 0.801 & 16.51 & 73.73 & 3.15 & 0.842 \\
&  & MGA-CLAP-audio
  & 10.97 & 58.89 & 2.90 & 0.800 & 16.50 & 73.53 & 3.14 & 0.841 \\
&  & M2D
  & 10.94 & 57.90 & 2.91 & 0.800 & 16.44 & 73.40 & 3.14 & 0.841 \\

\midrule

\rowcolor{gray!15} \cellcolor{white} \multirow[]{5}{*}{3} & M2D-SC & -
  & 11.00 & 55.80 & 2.88 & 0.802 & 14.94 & 61.80 & 3.06 & 0.835 \\
\cmidrule{2-11}
&  \multirow[]{4}{*}{DPC} & -
  & 11.06 & 58.64 & \textbf{2.92} & 0.802 & 16.34 & 72.93 & 3.13 & 0.839 \\
&  & MGA-CLAP-text
  & 11.18 & 59.38 & 2.91 & \textbf{0.803} & \textbf{16.60} & 74.20 & \textbf{3.16} & 0.842 \\
&  & MGA-CLAP-audio
  & 11.11 & 59.01 & 2.91 & 0.802 & 16.56 & \textbf{74.33} & 3.15 & 0.842 \\
&  & M2D
  & \textbf{11.19} & \textbf{59.63} & \textbf{2.92} & 0.802 & \textbf{16.60} & 74.20 & \textbf{3.16} & \textbf{0.843} \\

\bottomrule
\end{tabular}
}
\end{table*}

\section{EXPERIMENTS}
\label{sec:experiment}

\subsection{Datasets \& Evaluation metrics}
\label{subsec:dataset}
The training data was generated dynamically for each sample. We synthesized 4-channel mixtures on-the-fly using the SpatialScaper synthesis approach~\cite{10446118}. Additional data was incorporated for training. The speech class was supplemented with the VCTK corpus~\cite{veaux2017vctk}, and the percussion class was enriched with extra samples from open-source databases~\cite{techreport}. For evaluation, we reported results on the official DCASE 2025 task4 test and evaluation datasets. The test dataset was provided for the intermediate assessment during the challenge, while the evaluation dataset was used for the final official ranking of the challenge submissions. The 1,620-sample subset of the evaluation data was used , which is for the official challenge ranking.

We evaluated the performance of our systems using four metrics. The Class-Aware Signal-to-Distortion Ratio improvement (CA-SDRi)~\cite{yasuda2025, nguyen2025} is the primary metric. It is based on the standard SDRi, a measure of separation quality, but is designed to penalize incorrect classifications. If a source is misclassified, its SDRi is disregarded. The mixture-level accuracy $(\mathrm{Acc}_{\mathrm{mix}})$~\cite{dcaseworkshop} is used to evaluate classification performace. $\mathrm{Acc}_{\mathrm{mix}}$ is calculated on a per-mixture basis, where a single mixture is considered correct only if the classifications for all of its constituent foregrounds are correct. The Perceptual Evaluation of Speech Quality (PESQ)~\cite{Rix2001PerceptualEO} was used for the perceptual assessment of speech data. The PESQ score was calculated only for samples that were correctly classified as speech. For all non-speech classes, we used the CLAP-score to evaluate the semantic similarity between the separated audio and the target audio. This score is the cosine similarity between the LAION-CLAP~\cite{10095969} embeddings for correctly classified samples.

\subsection{Experimental settings}
\label{subsec:settings}
Our experiments were based on the pretrained weights of the previous system~\cite{techreport} on the challenge. To isolate the performance gains of our contributions, we froze the parameters of the core separation models (DM-USS and DM-TSE backbone). The only trainable modules in our experiments were the proposed DPC and SCE. The DPC was trained only once with class clues conditioned by one-hot vectors. After training the DPC, we froze its parameters and then trained the SCE separately using three distinct semantic embeddings: those from M2D and from the text and audio encoders of MGA-CLAP. AdamW optimizer was used for training both DPC and SCE. The learning rates for DPC and SCE were $4\times10^{-4}$ and $4\times10^{-6}$, respectively. Cross-entropy and KL divergence loss functions were used to train the classification ability of the DPC for active foreground and silence, respectively. Binary cross-entropy was utilized to detect silence. The masked SNR loss function was used for training the SCE. The detailed settings for the loss functions follow~\cite {dcaseworkshop}.

\section{Results}
\label{sec:results}

\subsection{Main performance comparison}
Table~\ref{tab:main} presents our main experimental results, organized by the framework's three stages. For each stage, we compare different models and clue types, reporting performance on both the evaluation and test datasets. We consider the evaluation dataset a more robust indicator of generalization, while Stage 3 represents the system's final, iteratively refined performance.

Regarding the classifier performance, the proposed DPC consistently outperforms the conventional M2D-SC in $\mathrm{Acc}_{\mathrm{mix}}$ in Stage 2 and 3 across all datasets. The improvements are particularly substantial on the test dataset, where DPC achieves accuracy gains of over 10\%p in every stage. Specifically, the accuracy increases from 59.07\% to 69.20\% (+10.13\%p) in Stage 1, from 61.26\% to 72.13\% (+10.87\%p) in Stage 2, and from 61.80\% to 72.93\% (+11.13\%p) in Stage 3. Notably, the margin of improvement for DPC over M2D-SC widens as the separation quality improves in the later stages. This suggests that DPC, by directly utilizing the object feature of the separators, is more sensitive and responsive to enhancements in the separators' performance. Furthermore, the DPC also demonstrates strong generalization. On the evaluation dataset, it achieves a 2.84\%p accuracy improvement in Stage 3. We attribute this generalization to the absence of fine-tuning on the M2D, which allows our systems to retain diverse features from its pretraining.

To evaluate the effectiveness of the SCE, the DPC is used for classification. The results in Table~\ref{tab:main} clearly demonstrate that incorporating any of the three semantic embeddings consistently improves performance over conditioning only on the one-hot vectors, particularly for the CA-SDRi. For instance, in Stage 3 on the test dataset, using the semantic embedding of M2D boosts the CA-SDRi from 16.34 dB to 16.60 dB (+0.26 dB). The improvement in the evaluation dataset from 11.06 dB to 11.19 dB (+0.13 dB) represents that the benefits of the SCE generalize well to unseen data.
While the three embeddings yield a similar level of improvement, the M2D embeddings demonstrate a slight edge in perceptual and semantic fidelity, achieving the best PESQ of 3.16 and the highest CLAP-score of 0.843 in Stage 3 on the test dataset.
Ultimately, these results indicate that the rich semantic clues provided by the SCE not only enhance the objective separation and classification quality (CA-SDRi) but also holistically improve the system's output, leading to more accurate classifications and audio that is perceptually and semantically more faithful to the target source.


\subsection{Classification error correction by SCE}

The error propagation to the next stage is a key challenge in the multi-stage framework. We therefore analyzed the error correction capabilities of the SCE to determine if it could mitigate this issue, with the results presented in Table~\ref{tab:ablation}. The table categorizes samples based on their transition of classification correctness from Stage 1 to Stage 2: Wrong-to-Correct (W2C) for corrected errors, Correct-to-Correct (C2C) for stable predictions, and Correct-to-Wrong (C2W) for newly introduced errors. For each category, we report the ratio of samples relative to the entire dataset and the average improvement in separation quality ($\Delta$CA-SDRi).

On the test dataset, using the SCE leads to comprehensive improvements across all categories for both the sample ratios and the $\Delta$CA-SDRi. This suggests that the SCE not only mitigates error propagation by improving classification transitions but also enhances the separation performance. Upon closer examination, different semantic embeddings exhibit distinct strengths. The M2D embedding proves to be the most effective at error correction, achieving the highest W2C ratio of 11.4\%. Meanwhile, the MGA-CLAP-text embedding excels at preserving correct predictions from the previous stage, evidenced by the highest C2C ratio of 62.4\% and the lowest C2W ratio of 6.8\%. Furthermore, the MGA-CLAP-audio shows the largest $\Delta$CA-SDRi of 11.86 dB for W2C and the smallest degradation of -5.27 dB for C2W, thereby boosting the overall performance.

\begin{table}[t]
\caption{Comparison of error correction performance by embedding type.}
\label{tab:ablation}
\centering
\setlength{\tabcolsep}{2pt}
{\scriptsize
\begin{tabular}{@{}cl*{6}{C{0.75cm}}@{}}
\toprule
\multirow[]{2}{*}{Dataset} & \multirow[]{2}{*}{Embedding type} & \multicolumn{3}{c}{Ratio of samples (\%)} & \multicolumn{3}{c}{$\Delta$ CA-SDRi (dB)} \\
\cmidrule(lr){3-5} \cmidrule(lr){6-8}
& & W2C $\uparrow$ & C2C $\uparrow$ & C2W $\downarrow$ & W2C $\uparrow$ & C2C $\uparrow$ & C2W $\uparrow$ \\

\midrule

\multirow[]{4}{*}{Test} & one-hot vector
  & 10.5 & 61.7 & 7.5 & 11.75 & 3.46 & -5.71 \\
& \hspace{0.5em} + MGA-CLAP-text
  & 11.3 & \textbf{62.4} & \textbf{6.8} & 11.78 & \textbf{3.67} & -5.63 \\
& \hspace{0.5em} + MGA-CLAP-audio
  & 11.3 & 62.3 & 6.9 & \textbf{11.86} & 3.64 & \textbf{-5.27} \\
& \hspace{0.5em} + M2D 
  & \textbf{11.4} & 62.0 & 7.2 & 11.82 & 3.64 & -5.71 \\

\midrule

\multirow[]{4}{*}{Eval} & one-hot vector
  & 13.5 & \textbf{44.3} & \textbf{7.8} & 9.89 & 2.43 & -4.60 \\
& \hspace{0.5em} + MGA-CLAP-text
  & 14.5 & 43.8 & 8.3 & \textbf{10.06} & 2.52 & -4.69 \\
& \hspace{0.5em} + MGA-CLAP-audio
  & \textbf{14.8} & 44.1 & 7.9 & 10.04 & 2.51 & -4.83 \\
& \hspace{0.5em} + M2D 
  & 14.4 & 43.5 & 8.6 & 10.01 & \textbf{2.53} & \textbf{-4.48} \\

\bottomrule
\end{tabular}
}
\end{table}

On the evaluation dataset, using only the one-hot vector shows slightly superior prediction preservation in terms of sample ratio. For example, it outperforms the second-best MGA-CLAP-audio embedding in the C2C ratio, 44.3\% to 44.1\%. Despite this, the benefit of the SCE becomes clear when observing the $\Delta$CA-SDRi. For W2C samples, all semantic embeddings outperform the one-hot vector by approximately more than 0.1 dB. For the other categories, the M2D embedding demonstrates an improvement of 0.10 dB from 2.43 dB to 2.53 dB on the C2C and 0.12 dB from -4.60 dB to -4.48 dB on the C2W. Collectively, this analysis confirms that the use of the SCE with additional semantic embeddings improves the final CA-SDRi performance. It achieves this effectively by mitigating error propagation or enhancing the quality of separation.

\section{Conclusion}
\label{sec:conclusion}

In this paper, we addressed two key limitations of the first-place system from DCASE 2025 Task 4: an inefficiently integrated classifier architecture and a weak conditioning mechanism susceptible to error propagation. To resolve these issues, we introduced the Dual-Path Classifier (DPC), which operates on object features from the separator and semantic features from a pretrained model. We also proposed the Semantic Clue Encoder (SCE) to enrich the conditioning signal with content-based embeddings from large-scale models. Our experiments demonstrate that the DPC significantly outperforms the baseline classifier and that the SCE provides substantial additional gains, achieving a state-of-the-art CA-SDRi of 11.19 dB on the evaluation set. Furthermore, our analysis revealed that the SCE enhances system robustness through a dual mechanism: it not only corrects a greater number of initial classification errors (W2C), but also improves the separation quality with the rich semantic information.


\pagebreak

\section{acknowledgements}
This work was supported by the National Research Foundation of Korea (NRF) grant (No. RS-2024-00337945) and STEAM research grant (No. RS-2024-00464269) funded by the Ministry of Science and ICT of Korea government (MSIT), the BK21 FOUR program through the NRF grant funded by the Ministry of Education of Korea government (MOE).
\begingroup
\footnotesize
\bibliographystyle{IEEEbib}
\bibliography{strings,refs}
\endgroup

\end{document}